\documentclass{article}
\usepackage[T1]{fontenc} 
\usepackage[utf8]{inputenc} 
\usepackage{ismir,amsmath,cite,url}
\usepackage{graphicx}
\usepackage{color}

\def\code#1{\texttt{#1}}
\title{Leveraging Hierarchical Structures for \\ Few-Shot Musical Instrument Recognition}

\oneauthor
{Hugo Flores Garcia, Aldo Aguilar, Ethan Manilow, Bryan Pardo}
{ {\normalsize \tt \{hugofg@u., aldoa@u., ethanm@u., pardo@\}northwestern.edu} \\Interactive Audio Lab, Northwestern University, Evanston, IL, USA}

\def\authorname{H. Flores Garcia, A. Aguilar, E. Manilow and B. Pardo}

\begin{document}

\maketitle

\begin{abstract}

Deep learning work on musical instrument recognition has generally focused on instrument classes for which we have abundant data. In this work, we exploit hierarchical relationships between instruments in a few-shot learning setup to enable classification of a wider set of musical instruments, given a few examples at inference. We apply a hierarchical loss function to the training of prototypical networks, combined with a method to aggregate prototypes hierarchically, mirroring the structure of a predefined musical instrument hierarchy. These extensions require no changes to the network architecture and new levels can be easily added or removed. Compared to a non-hierarchical few-shot baseline, our method leads to a significant increase in classification accuracy and significant decrease in mistake severity on instrument classes unseen in training.

\end{abstract}
\section{Introduction}\label{sec:introduction}


Musical instrument recognition is a machine learning task that aims to label audio recordings of musical instruments, typically at a fine temporal granularity (second by second)~\cite{fu2010survey,eronen2000musical,krishna2004music}. Musical instrument recognition can be viewed as a subtask of Sound Event Detection (SED), which consists of identifying and locating any type of sound event (\textit{e.g.}, car horn, dog bark) in an audio recording ~\cite{mesaros2016metrics,mesaros2016tut,salamon2017deep}.


Labelling audio tracks is extremely important for organizing the dozens of tracks in a typical Digital Audio Workstation (DAW) recording session~\cite{savage2011art,owsinski2013mixing}, but manual labelling is a tedious process. Automated musical instrument recognition could enable automated track labeling. Automated second-by-second labeling could go further, enabling navigation through recording projects by traversing musical instrument \textit{labels}, rather than waveform visualizations. This would be especially helpful for audio engineers with low or no vision, as existing interfaces leave accessibility as an afterthought \cite{saha-vision-2020} and navigating by visually examining waveforms is not a viable option for them \cite{tanaka2016haptic}.

A barrier to incorporating instrument recognition into DAWs is that most existing deep learning techniques must be trained on instruments that have abundant labeled training data. The datasets that support these systems only focus on the limited set of instrument classes that have sufficient data~\cite{bosch2012comparison,humphrey-openmic-2018,hung-timbre-2018,gururani-attention-2019,hung-multitask-2019,taenzer2019investigating, kratimenos2021augmentation}. However, the vast diversity of musical instrument sounds necessitates supporting a broader set of instrument classes~\cite{lostanlen2018extended}. While expanding current datasets with more diverse coverage 
can ameliorate this issue, collecting human annotations for a large number of audio files is a tedious, time consuming task \cite{kim-ised-2017, cartwright2019crowdsourcing}, and there will always be unanticipated sound categories that an end-user would like to automatically label.

\begin{figure}
    \centering
    \includegraphics[width=0.9\columnwidth]{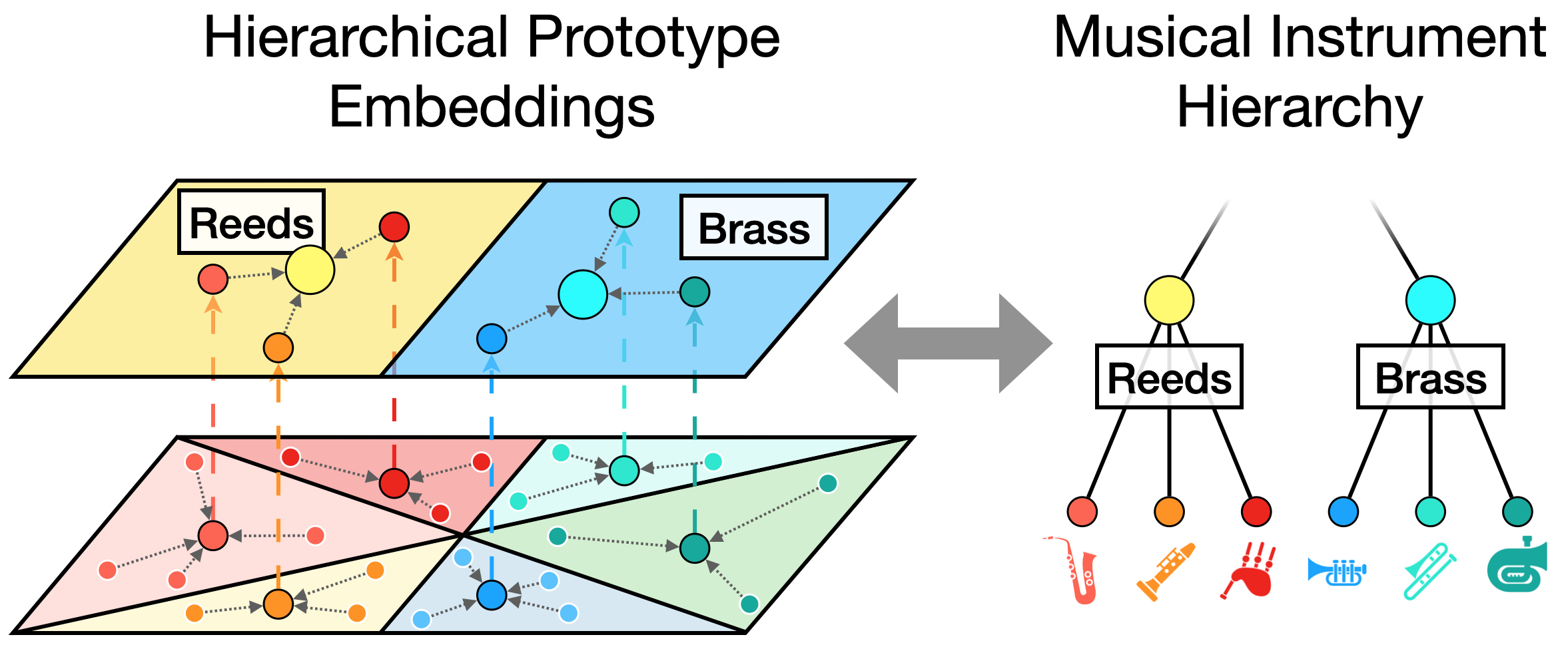}
    \caption{Overview of our method. Prototypes from a set of embedded support examples at a fine-grained level (bottom left) are aggregated to make a set of \textit{metaprototypes} at a coarser-grained level (top left). In this way, we learn a hierarchical set prototypes that corresponds to a musical instrument hierarchy (right).}
    \label{fig:hier_prototypes}
     \vspace{-12pt}
\end{figure}

Therefore, musical instrument recognition systems should be able to dynamically expand their vocabularies after deployment, to conform to end-user needs. This requires an approach that lets a system learn a new sound category given only a few examples that can be provided by an end user, \textit{a la} few-shot learning.


 Using a hierarchical system, like the widely-used Hornbostel-Sachs hierarchy~\cite{hornbostel-classification-1961}, to organize and classify musical instruments has broad precedent in many human cultures ~\cite{kartomi1990concepts}. 
 We can take advantage of a musical instrument hierarchy, like the widely-used Hornbostel-Sachs hierarchy~\cite{hornbostel-classification-1961}, to improve few-shot learning. A system could learn a feature space meaningful for unseen classes that share hierarchical ancestry with the classes seen during training. For example, the Chinese zhongruan is a plucked string instrument that shares ancestry with other chordophones in the Hornbostel-Sachs hierarchy (like the guitar), which might be more common in datasets of Western instruments. A model could leverage the hierarchical relationship between an instrument it has never been trained on (\textit{e.g.} the zhongruan) and more common instruments seen during training (\textit{e.g.} the guitar) to produce a meaningful representation of the new instrument with only a few support examples.

In this work, we propose a simple extension to prototypical networks \cite{snell-prototypical-2017} that imposes a hierarchical structure on the learned embedding space (Figure \ref{fig:hier_prototypes}). We first create prototypes from an initial set of embedded support examples at the most granular level. We then aggregate these initial prototypes into new prototypes corresponding to a coarser hierarchical level, in a manner reminiscent of agglomerative clustering~\cite{oded2006datamining}. Repeating this process lets our system represent classes at many granularities of a predefined instrument hierarchy. We also propose a weighted, hierarchical extension of cross-entropy loss to ensure the network learns the hierarchy. Compared to a non-hierarchical few-shot baseline~\cite{wang-fewshotsed-2020}, our method shows a significant increase in classification accuracy and significant decrease in mistake severity on unseen instrument classes.

 \vspace{-12pt}
\section{Related Work}\label{sec:related_work}
Musical instrument recognition can be performed in single-source contexts \cite{benetos-nmf-2006, eronen-mfcc-2000, lostanlen-spiral-2016, essid-hierarchicalsolos-2006}, where only a single sound source may be active at any given time, as well as in multi-source contexts \cite{han-predominant-2017,hung-multitask-2019,hung-timbre-2018, gururani-attention-2019,gururani-iad-2018}, where multiple sound sources may be active at the same time. We consider the single-source case, as the vast majority of audio in a studio music production workflow is single-source.

Hierarchical structures have shown to be effective for many machine learning tasks, such as text classification \cite{stein-hierarchicaltext-2019} and image classification \cite{ankit-image_hierarchical-2020, sun-hierarchicalimage-2019}.
In fact, Bertinetto \textit{et al.} \cite{bertinetto-mistakes-2020} propose a hierarchical image classification approach that uses a similar exponentially weighed hierarchical loss function to the one proposed here, although they do not focus on a few-shot setting, as we do, and they favor learning broader classes, whereas we are also interested in finer classes.
Hierarchical structure was explored for musical instrument recognition by using fixed signal processing feature extraction techniques~\cite{essid-hierarchicalsolos-2006, essid-2005-instrument, kitahara-hierarchicalnonregistered-2004}. Here, we use deep learning methods to flexibly learn a feature space that mirrors musical instrument hierarchies. 

Recent work has studied how hierarchical structures can be incorporated into neural network models for different tasks. In the automatic speech recognition (ASR) domain, CTC-based hierarchical ASR models \cite{fernandez-hierarchicalsequencernn-2007, sanabria-hierarchicalctc-2018, krishna-hierarchicalspeech-2018} employ hierarchical multitask learning techniques, particularly by using intermediate representations output by the model to perform intermediate predictions in a coarse-to-fine scheme. 
Manilow \textit{et al.} \cite{manilow-hierarchical-2020} have shown that hierarchical priors can have significant benefits for performing source separation of musical mixtures. None of these systems, however, were designed for few-shot learning. 

Previous deep learning systems have been proposed for multilevel audio classification~\cite{xu2016hierarchical,jati-hierarchical_loss-2019,cramer-taxonet-2020}. However, none of these systems work in a few-shot setting and they require either specialized network architectures or complex data pipelines to learn a hierarchy. Our approach is a simple extension to incorporate hierarchy into an established few-shot learning paradigm. 

Recent work in audio tagging and sound event detection tasks has explored few-shot learning in the audio domain~\cite{kim-ised-2017,cheng-fewshotsed-2019,shi-fewshotacoustic-2020,wang-fewshotdrum-2020,wang-fewshotsed-2020}, though none of this work assumed any hierarchical structure. 


Here, we propose a method for hierarchical representation learning in a few-shot setting, leveraging the increased flexibility of both hierarchy and few-shot methods for musical instrument recognition.

\vspace{-0.5cm}

\section{Background}

\subsection{Few-shot Learning}
\label{sec:few_shot}

In a few-shot classification setting, we consider a target class $k \in \mathcal{K}$ for a set of target classes,  $\mathcal{K}$, of size $|\mathcal{K}|$. 
Let $x_s$ be a single support example drawn from a set of examples $\mathcal{S}$, called the support set. Assume $N$ labeled support examples (\textit{i.e.}, shots) per class $k$, totalling $N \times |\mathcal{K}|$ labeled examples. We define $\mathcal{S}_k$ as the subset of $\mathcal{S}$ containing the examples of class $k$.

We are provided an unlabeled query set $\mathcal{Q}$ of $M$ unlabeled examples. The goal of the task is to label each query example $x_q \in \mathcal{Q}$ with a target class $k \in \mathcal{K}$. A neural network model $f_\theta$ projects both the support and query sets into a discriminative embedding space. The query is assigned to the class of the support set it is closest to, according to distance metric $d$. 

\subsection{Prototypical Networks}\label{sec:protonets}

Prototypical networks~\cite{snell-prototypical-2017} compute an embedding vector for each instance in $\mathcal{S}_k$.
The prototype, $c_k$, for class $k$ is the mean vector of all the support embeddings belonging to class $k$:

\begin{equation} \label{eq:proto}
    c_k = \frac{1}{|S_k|} \sum_{x_s \in \mathcal{S}_k} f_\theta(x_s).
\end{equation}

Using a distance function $d$, we can produce a probability distribution over the set of classes $\mathcal{K}$ for a given query $x_q$ by applying a softmax over the negated distances from the query to each class prototype:

\begin{equation} \label{eq:proto-softmax}
    p(\hat{y}_{q}=k|x_q) = \
        \frac{\exp{(-d(f_\theta(x_q), c_{k}))}}
             {\sum_{c'_{k}} \exp{(-d(f_\theta(x_q), c'_{k}))}}.
\end{equation}
We use the Euclidean distance as $d$ in this work.

\section{Method}

Musicologists have long categorized musical instruments into hierarchical taxonomies, such as the Hornbostel-Sachs system \cite{hornbostel-classification-1961}, which classifies musical instruments into a hierarchy corresponding to their sound producing mechanisms. We can improve upon existing few-shot models by leveraging the hierarchical structure intrinsic to musical instrument taxonomies. To do this, we extend prototypical networks by training on a multitask scenario composed of multiple classification tasks, one for each level of a class tree, where the prototype for a parent node in the class tree is defined as the mean of the prototypes for each of the parent node's children.

We impose hierarchical structure on our few-shot task by constructing a tree, $T$, with height $H$, starting from a set of leaf nodes. We define the leaf nodes as the same set of classes, $\mathcal{K}$, that we defined for our standard few-shot setup in Sec.~\ref{sec:few_shot}. We then define the parents of the leaf nodes by aggregating classes, $k \in \mathcal{K}$. For musical instrument recognition, we aggregate classes according to a predefined instrument hierarchy (\textit{e.g.}, Hornbostel-Sachs). We iteratively aggregate child classes up to the max height of the tree $H$. We index the tree as $T_{i, h}$, where $i \in \mathcal{K}_i$ indexes over the set of sibling classes at level $h$, for $h=0, \dots H$, with level $0$ containing the most specific classes and level $H$ containing the broadest. In our notation $H=0$ describes a tree with no hierarchy and is equivalent the non-hierarchical prototypical network defined in Sec.~~\ref{sec:protonets}. $H=1$ has two levels, and so on.


\subsection{Hierarchical Prototypical Networks}
\label{sec:hproto}

We define our proposed hierarchical prototypical network by extending typical prototypical networks~\cite{snell-prototypical-2017} to a hierarchical multitask learning scenario, where we wish to label each query example, $x_q \in \mathcal{Q}$, at multiple levels of our class tree, $T$. Here, labeling at each level is a separate task. 

Like a normal prototypical network, we use a network $f_\theta$ to produce embeddings for every example in the support set. The mean of these embedded support examples creates an initial set of prototypes (Eq.~\ref{eq:proto}). We deviate from the typical setup by considering this initial set of prototypes as the lowest level of our tree, $T$, and aggregating these initial prototypes \textit{again} to make another set of prototypes representing the next level. The prototypes at this higher level are, thus, prototypes of prototypes, or \textit{metaprototypes}, and define a hierarchy according to the structure of our tree, $T$. We continue to iteratively aggregate prototypes in this fashion for all levels of our tree.  The prototype for each parent class at level $h+1$ is notated $c_{T_{i, h+1}}$ and is the mean of the members of its support set $\mathcal{S}_{T_{i, h}}$. For levels $h>0$, each example $\hat{x}_s$, is itself a prototype:

\begin{equation} \label{eq:hproto}
c_{T_{i, h+1}} = \frac{1}{|\mathcal{S}_{T_{i, h}}|} \sum_{\hat{x}_s \in \mathcal{S}_{T_{i, h}}} f_\theta (\hat{x}_s) ,
\end{equation}

This process is shown in Figure \ref{fig:hier_prototypes}.

Given a query example $x_{q}$, we use the network to create an embedding $f_\theta(x_q)$ and measure its distance to each class prototype or metaprototype $c_{T_{i, h}}$ at a given level $h$. Given these distances, we output $H$ probability distributions, one for each level in our class tree:

\begin{equation} \label{eq:hproto-softmax}
    p(T_{i, h}|x_q) = \
        \frac{\exp{(-d(f_\theta(x_q), c_{T_{i, h}}))}}
             {\sum_{c'_{T_{i, h}}} \exp{(-d(f_\theta(x_q), c'_{T_{i, h}}}))}.
\end{equation}

We note that Eqs.~\ref{eq:proto} and \ref{eq:proto-softmax} are special cases of the proposed Eqs.~\ref{eq:hproto} and \ref{eq:hproto-softmax}, evaluated at $h = 0$.
Our generalization allows multi-task few-shot classification at multiple levels of a hierarchical class tree.

Our proposed method does not require any specific network architecture. 
Instead, it provides a hierarchical label structure for support examples $x_s$ to be aggregated together, forming fine-to-coarse representations (\textit{i.e.}, $c_{T_{i, h}}$) that we can leverage and optimize with. This exposes the potential for a model to be trained with multiple concurrent hierarchies, a direction for future work.    

\subsection{Multi-Task Hierarchical Loss} \label{sec:method-loss}

We now set up a learning objective, where we minimize the cross-entropy loss between the predicted distribution and the ground truth class for each level in the class tree. The intuition behind our approach is that we can use a hierarchically structured objective to encourage our model to produce an embedding space with discriminative properties at both coarse and fine granularities, allowing some of these coarse features to generalize beyond the training set of fine grained leaf classes to their unseen siblings in the class tree. We use an exponentially decaying sum of loss terms for each level in the hierarchy~\cite{bertinetto-mistakes-2020}:

\begin{equation}
    \mathcal{L}_{hierarchical} = \sum_{h=0}^{H} e^{-\alpha \cdot h} \mathcal {L}_{CE}^{(h)},
\end{equation}

where $\mathcal {L}_{CE}^{(h)}$ denotes the cross-entropy loss for the classification task at height $h$, and $\alpha$ is a hyperparameter that determines the decay of each loss term w.r.t height. Setting $\alpha > 0$ places more more weight on finer-grained tasks, $\alpha < 0$ places more weight on coarser-grained tasks, and $\alpha = 0 $ weighs all tasks equally. 
We note that $H = 0$ reduces to the non-hierarchical (baseline) definition of the problem, where we only optimize for the fine-grained task. 



\section{Experimental Design}
\label{sec:exp_design}
We evaluated our proposed hierarchical prototypical approach using a non-hierarchical prototypical method~\cite{wang-fewshotsed-2020} as a baseline. We evaluated all models on a few-shot musical instrument recognition task, measuring standard classification metrics (F1) as well as mistake severity. We conducted ablations for class tree height, choice of class hierarchy, and proposed loss function. 


\subsection{Datasets}
For all experiments, we trained and evaluated using isolated tracks from the MedleyDB\cite{bittner-medleydb-2014} and MedleyDB 2.0 \cite{bittner-medleydb2-2016} datasets. MedleyDB contains multi-track recordings of musical instruments and vocals. We excluded recordings that do not have fine-grained instrument labels (\textit{e.g.}, ``brass'' was excluded because the audio could be of trumpets, trombones, etc.). Additionally, we considered sections of a single instrument to be the same class as the instrument itself (\textit{e.g.} "violin section" and "violin" both belong to the class "violin"). Altogether, the dataset consists of 63 different instruments, with 790 tracks in total.

For training and evaluation, we removed the silent regions of each audio track. We then split the remainder of the track into 1 second segments with a hop size of 0.5 seconds, where each 1 second segment is an input example to the model. All audio was downsampled to 16kHz. For each example, we compute a 128-bin log-Mel spectrogram with a 32ms window and an 8ms hop. 
After preprocessing, our training and evaluation datasets contained 539k and 56k 1-second examples, respectively. 
We performed silence removal using \code{pysox} \cite{bittner-pysox-2016}.

\subsection{Network Architecture}
The backbone network architecture used in all experiments was based on the prototypical network described in Wang \textit{et al.} \cite{wang-fewshotdrum-2020}. It uses a log-Mel spectrogram as input, and consists of four CNN blocks, where each convolutional filter has a kernel size of $3 \times 3$, followed by a batch normalization layer, a ReLU activation, and a $2 \times 2$ maxpooling layer. After the last convolutional block, we applied maxpooling over the time dimension, to obtain a 1024-dimensional embedding. Finally, we added a linear projection layer that reduces the 1024-dimensional embedding to 128 dimensions. 

\subsection{Hornbostel-Sachs Class Tree}\label{sec:classtree}
We used a musical instrument hierarchy inspired by the Hornbostel-Sachs~\cite{hornbostel-classification-1961} taxonomy,\footnote{See: https://en.wikipedia.org/wiki/Hornbostel-Sachs} (maximum height of 4) which is organized by the sound production mechanisms of each instrument. Since similar sound production mechanisms can lead to similar sounds, we believe this is a natural organization that our model can leverage to learn discriminative features at different levels of a class hierarchy.
 
\subsection{Episodic Training and Evaluation} \label{sec:episodic-eval}

We have a musical instrument hierarchy tree, where individual instrument classes are leaf nodes (e.g. violin, guitar). Nodes at higher levels ($h>0$) are instrument families, (e.g. bowed strings, plucked strings). Our goal is to observe classification performance on previously-unseen leaf classes (e.g. zhongruan, erhu). Therefore, we created a data split of 70\% train, 30\% evaluation, with no overlap between train and evaluation classes at the leaf instrument level ($h=0$). We further added the constraint that the classes in both testing and evaluation sets be distributed evenly among the instrument families ($h>0$). This avoids a problem where, for example, the train set consists only of percussion and the evaluation set consists only of chordophones. All experiments shared a train/evaluation split.

For each experiment, we trained every model in a few-shot learning scenario using episodic training. Each model was presented with a unique $\mathcal{|K|}$--way, $N$--shot learning task (an episode) with $M$ queries per leaf class at each training step. We constructed an episode by sampling a set of $|\mathcal{K}|$ instrument classes from the training data. For each of these $|\mathcal{K}|$ classes, we sampled $N + M$ audio examples. Here, for each class $k$, $N=|S_k|$ is the number of "shots" in the support set and $M$ is the size of the query set. 

We trained all models using the same random initialization for a maximum of 60,000 steps with early stopping after the evaluation loss stopped improving for 4500 steps, using the Adam optimizer and a learning rate of 0.03. During training, we set $|\mathcal{K}| = 12$, $N = 4$, and  $M = 12$.
We evaluated each trained model on episodes constructed from the test data. For each evaluation, we made 100 episodes, with $|\mathcal{K}| = 12$, $M = 120$. All hyperparameters were fixed except those we ablated, as described below.

\subsection{Evaluation Metrics}
\label{sec:evaluation-metrics}

We used the F1-score as our primary classification metric, reporting the distribution of F1 scores computed for each episode, evaluated for predictions made at the finest level of the hierarchy. 

Similar to Bertinetto \textit{et al.} \cite{bertinetto-mistakes-2020}, we used the hierarchical distance of a mistake as a metric indicative of a model's mistake severity. Given a class tree, the hierarchical distance of mistake is defined as the height of the lowest common ancestor (LCA) between the prediction node and ground truth node when the input is misclassified (that is, when the model makes a mistake).
We report the average hierarchical distance of a mistake over all evaluation episodes.

For all hierarchical models, we measured mistake severity with respect to its own hierarchy. For the non-hierarchical model, we evaluated with respect to our proposed 4-level version of the Hornbostel-Sachs hierarchy, as we believe that its organization is meaningful.

\section{Experiments}
We now describe specific experiments to measure the effects of different design choices. We trained and evaluated all models using the procedure described in Section~\ref{sec:exp_design}. Our experiment code is available online~\footnote{https://github.com/hugofloresgarcia/music-trees}. 

\subsection{Tree Height}
To observe the effect of tree height on classification, we constructed shorter trees from the Hornbostel-Sachs class tree by removing every leaf node's parent until the desired max height of the tree is met. We trained and evaluated five models using our proposed class tree, shortened to different heights $H \in \{0, 1, 2, 3, 4\}$, where $H = 0$ is the baseline, non-hierarchical case inspired by Wang \textit{et al}. \cite{wang-fewshotsed-2020}. Each model was trained with $\alpha = 1$ and evaluated with $N = 8$ support examples per class, at inference.

\begin{figure}
 \centerline{
 \includegraphics[width=1\columnwidth]{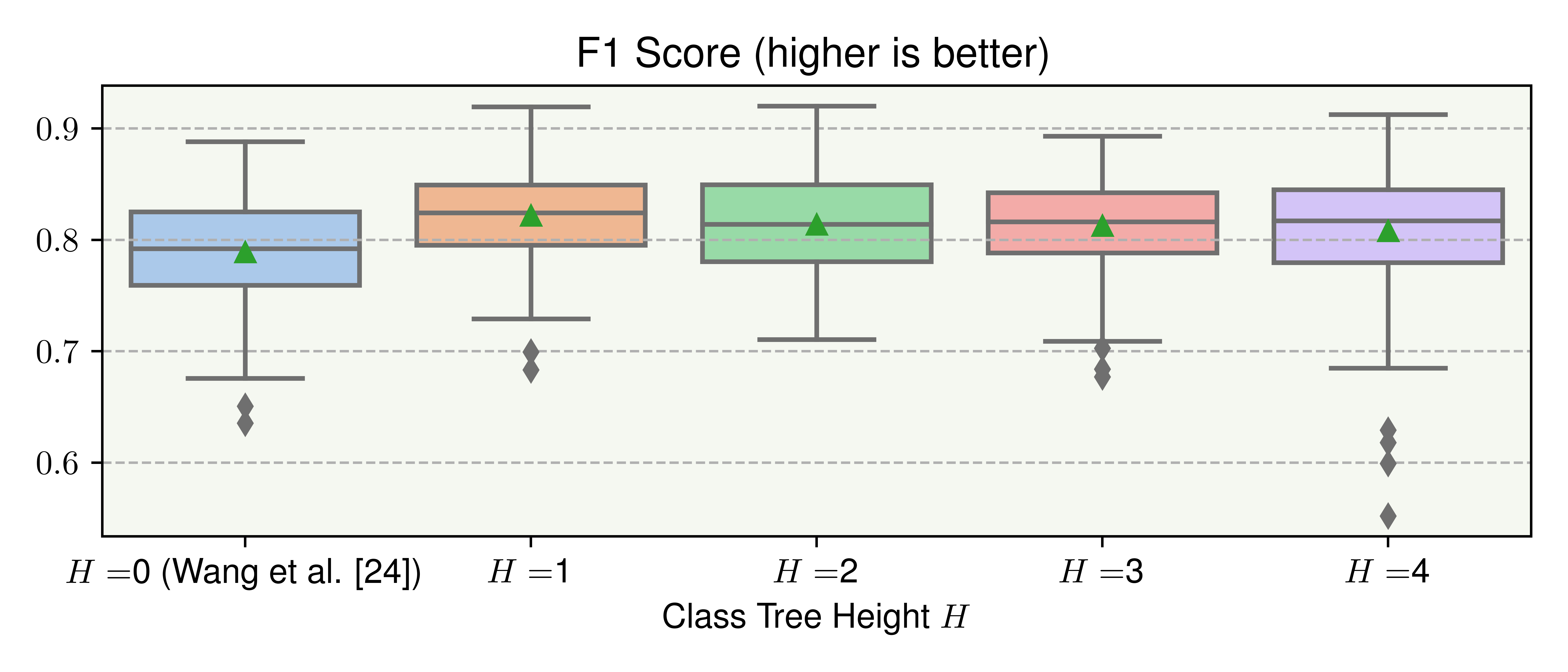}
 \vspace{-12pt}}
 \caption{F1 scores for models trained with class trees of varying height $H$, evaluated over 100 episodes. Means are shown as green triangles. Note that $H = 0$ is our baseline model (Wang \textit{et al}.~\cite{wang-fewshotsed-2020}), as it is trained without a class tree.}
 \vspace{-12pt}
 \label{fig:height}
\end{figure}

Results are shown in Figure \ref{fig:height}.  All variations of the proposed model achieved a better classification performance than the baseline. The best F1 score was seen at $H = 1$, with a mean value of .8111 over all evaluation episodes. Compared to the baseline mean score of .7792, this is a 4\% improvement.  A Wilcoxon signed-rank test showed that all of our proposed models achieve a statistically significant improvement when compared to the baseline, with $p < 10^{-7}$ for all hierarchies. These results show that incorporating our method into a prototypical network can lead to statistically significant improvements in classification performance under few-shot learning conditions. 

Surprisingly, a shallow tree with only the coarsest categories and the leaf nodes ($H = 1$) achieved the highest increase in performance. We believe this is due to the small number of classes encountered in a training episode (in our case, 12).
At a given level of the tree, at least 2 of the classes in the support set need to have a parent node in common for our method to be able to compute a meaningful metaprototype that can be leveraged by our loss. As a class tree gets deeper, the number of nodes at a given level can grow exponentially, meaning that our support set of 12 classes has a lower chance of finding meaningful groupings at deeper levels. This indicates that loss terms for levels closer to the leaf nodes are more likely be identical to the non-hierarchical loss.
Though the loss term for the coarsest level is still present in these deeper trees, it has a smaller impact on the gradient of the primary loss function, as loss terms are weighted to decay exponentially as the height increases. We believe training with a higher $|\mathcal{K}|$ can help leverage deeper hierarchies better. However, we leave this for future work. 

\subsection{Number of Support Examples}

\begin{figure}
 \centerline{
 \includegraphics[width=1\columnwidth]{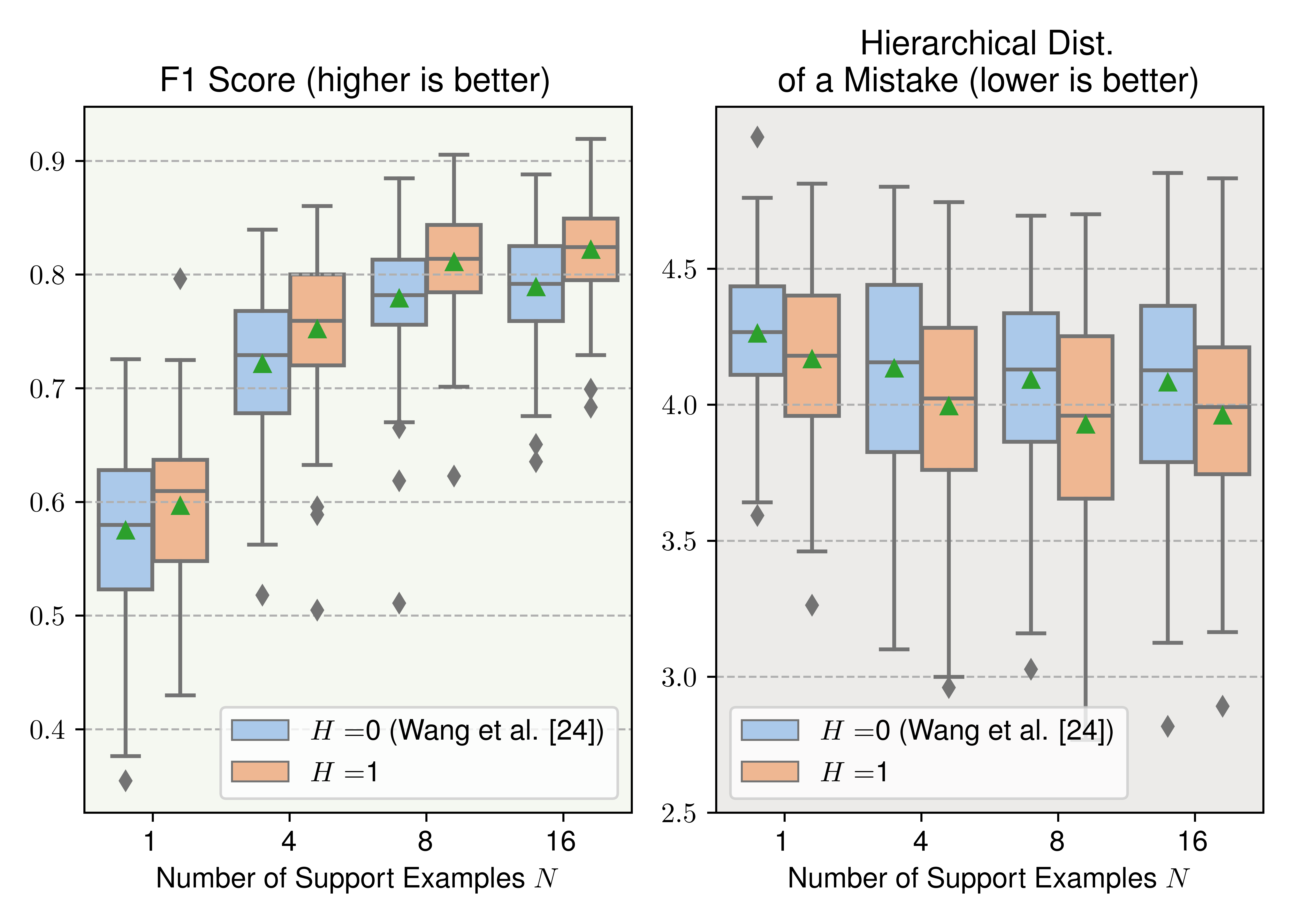}}
 \caption{Model comparison between the baseline model and our best proposed model  ($H = 1$), evaluated under conditions with a different number of shots (support examples) provided during inference. }
 \vspace{-12pt}
 \label{fig:nshot}
\end{figure}

We evaluated our best proposed model ($H = 1$, $\alpha = 1$) as well as our baseline model by varying the number of support examples $N$ provided to the model, where $N \in \{1, 4, 8, 16\}$. Results are shown in Figure \ref{fig:nshot} (left). We notice that increases in performance are greater when more support examples are provided, with the smallest increase (+2.17\% in the mean relative to baseline) occurring when $N = 1$. Our model achieved a statistically significant improvement on all test cases ($p < 10^{-4}$ for all $N$). 

As shown in Figure \ref{fig:nshot} (right), our model achieved a lower hierarchical distance of a mistake, on average. A Wilcoxon signed-rank test indicates that all improvements are statistically significant ($p < .0005$). This means that, when making incorrect predictions, our method was more likely to make predictions that are closer to the ground truth in terms of the class hierarchy (\textit{i.e.}, lower mistake severity). We believe it is fair to assume that mistake severity from a sound production perspective (as in our class hierarchy) is related to mistake severity in predictions made by humans. That is, a human is more likely to confuse a viola for a violin than to confuse a viola for a drum. 

\begin{figure*}[t]
    \centering
    \includegraphics[width=\textwidth]{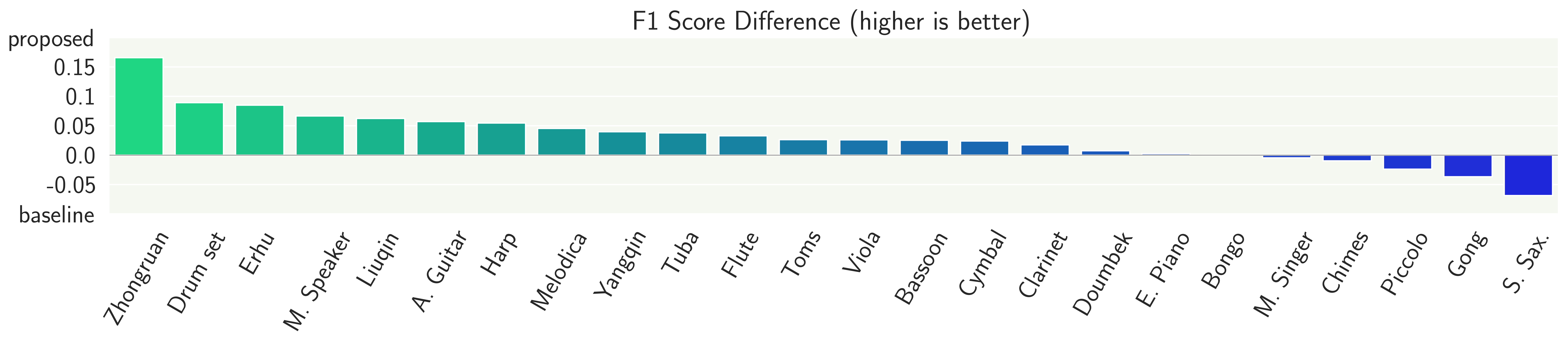}
    \vspace{-14pt}
    \caption{Difference in F1-score between our best proposed model  ($H=1$, $\alpha=1$) and the baseline (Wang \textit{et al.} \cite{wang-fewshotsed-2020}) on all instruments in the test set. Both models were evaluated with $N = 8$.}
    \vspace{-12pt}
    \label{fig:all_insts}
\end{figure*}

\vspace{-6pt}
\subsection{Arbitrary Class Trees}

To understand how the choice of hierarchy affects the results of our model, we evaluated the same prototypical network architecture trained using the Hornbostel-Sachs hierarchy and also 10 randomly generated class trees. We generated each tree by performing random pairwise swaps between leaf nodes in our original class tree, doing so 1000 times for each node. For this experiment, all trees were trained with ($H = 3$, $\alpha=1$), and evaluated with $N = 16$. 

\begin{figure}
 \centerline{
 \includegraphics[width=1\columnwidth]{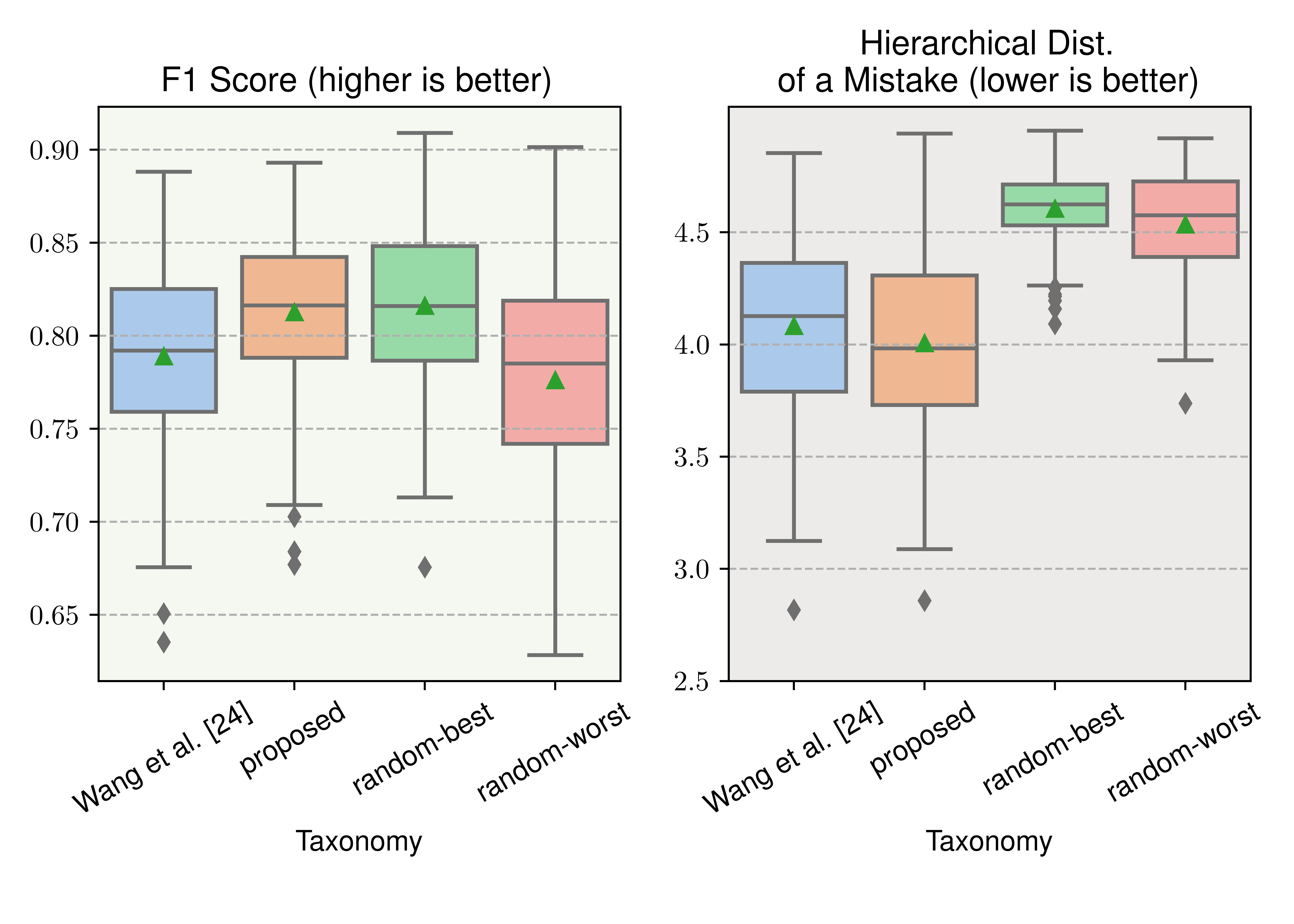}}
  \vspace{-12pt}
 \caption{Comparison between the best and worst performing models trained on random hierarchies. The hierarchical distance of a mistake is calculated using the hierarchy the model was trained on. For the baseline (Wang \textit{et al.}~\cite{wang-fewshotsed-2020}) , we calculated the hierarchical distance of a mistake using the Hornbostel-Sachs hierarchy.}
 \vspace{-12pt}

 \label{fig:random}
\end{figure}

Results for our evaluation of random class hierarchies are shown in Figure \ref{fig:random}. Our best performing random hierarchy in terms of classification performance ("random-best") achieves an F1 score comparable to our proposed hierarchy ($p > 0.05)$ though with a larger spread. Additionally, "random-best" obtains much worse mistake severity relative to the hierarchy it was trained on. This indicates that the model was not able to generalize the hierarchical structure it was trained on to out-of-distribution classes. On the other hand, our worst performing random hierarchy, "random-worst", caused a statistically significant deterioration in both classification performance and mistake severity compared to the baseline ($p < 0.005$). Even though the random-best model fairs comparably to Hornbostel-Sachs model, it is impossible to know \textit{a priori} whether any random tree will produce good results, therefore for practical uses (\textit{i.e.}, within a DAW), we find Hornbostel-Sachs to be a suitable choice.

\subsection{Hierarchical Loss Functions} \label{sec:loss-ablation}

To measure the impact of our proposed multi-task hierarchical loss, we compared it to a reasonable baseline "flat" loss. 
As our baseline approach, we treated hierarchical classification as a single-task, multilabel classification problem, where the ground truth is a multi-hot vector, with $1$s for the leaf ground truth node and all of its ancestors in the tree, and $0$s otherwise. Furthermore, we minimized the binary cross entropy between each individual predicted node and ground truth node. Note that this required us to use a sigmoid function instead of Eq.~\ref{eq:proto-softmax}, which uses a softmax function. Additionally, we performed a hyperparameter search to find the best value of the $\alpha$ parameter for our proposed loss function (Section \ref{sec:loss-ablation}) using the search space $\alpha \in \{-1, -0.5, 0, 0.5, 1\} $. For this experiment, all trees were trained with $H = 4$ and evaluated with $N = 16$.

\begin{figure}
 \centerline{
 \includegraphics[width=1\columnwidth]{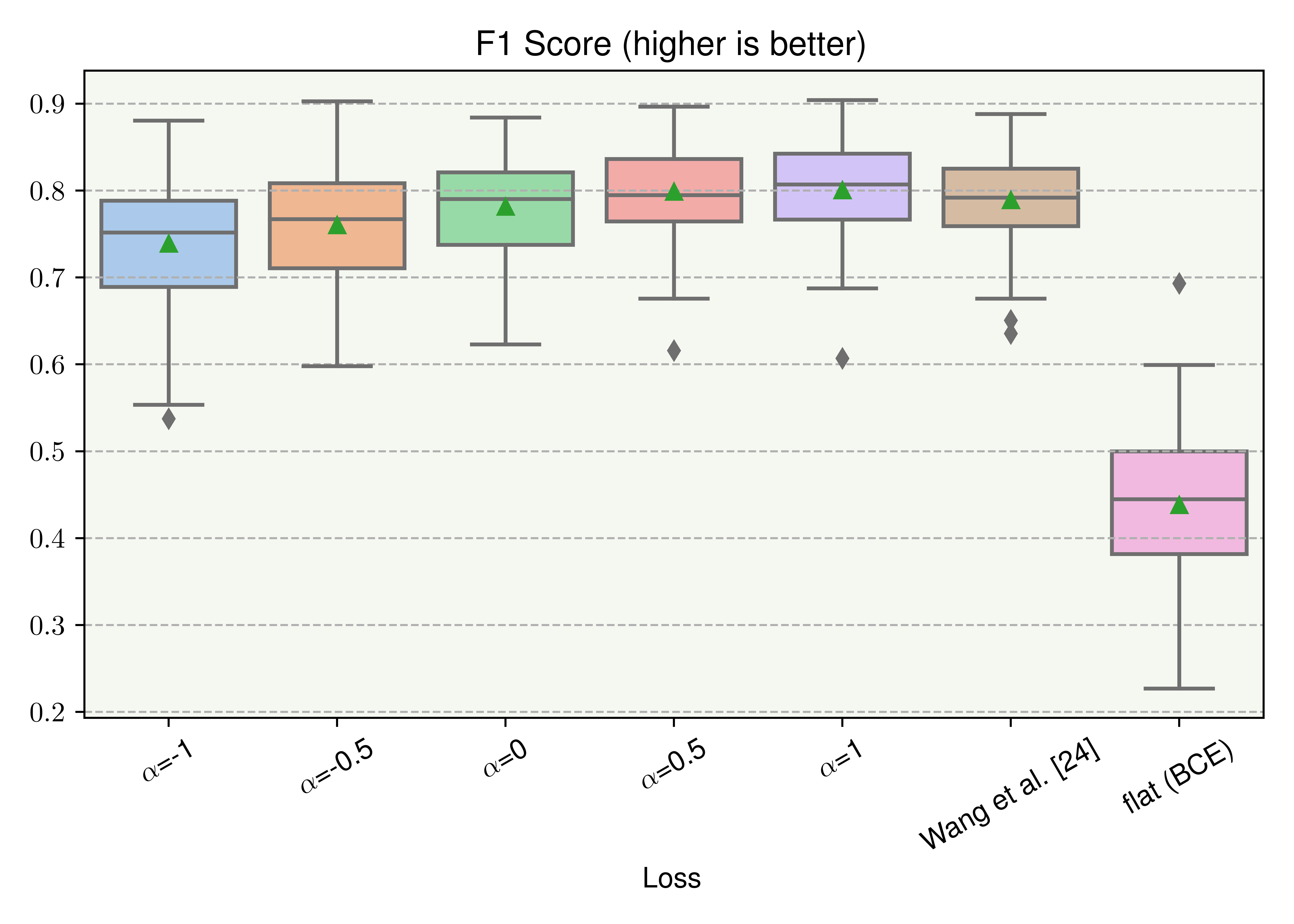}}
  \vspace{-12pt}
 \caption{Evaluating the loss function. We vary $\alpha$ in our proposed hierarchical loss from negative (emphasize loss on broader categories) to positive (emphasize loss on finer categories) and additionally compare to a "flat" binary cross entropy (BCE) baseline.  }
 \vspace{-12pt}
 \label{fig:alpha}
\end{figure}

Results are shown in Figure \ref{fig:alpha}. 
We observe that only the models with $\alpha > 0$ cause an improvement over Wang \textit{et al}. \cite{wang-fewshotsed-2020}. 
Moreover, the flat loss causes a severe degradation in classification performance. This may be because training prototypical networks using a binary, one-vs-all formulation could yield a much less discriminative embedding space. Wang et al. \cite{wang-fewshotsed-2020} found a similar result: training prototypical networks with a binary formulation did not yield performance improvements.

\vspace{-0.25cm}

\subsection{Examining All Instrument Classes}

In Figure~\ref{fig:all_insts}, we examine the classification performance of every instrument in our test set. We compare our best model ($H=1$, $\alpha=1$) to the baseline model from Wang \textit{et al.}~\cite{wang-fewshotsed-2020}, evaluated with $N = 8$. For clarity, we report the difference in F1 Score between the models. Our model beats the baseline on 18 of the 24 classes in the test set. In particular, our model shows a substantial improvement ($+16.56\%$) in F1 Score when classifying \textit{zhongruan}, which may be rarely seen in a dataset composed of Western music. Figure~\ref{fig:all_insts} demonstrates that, overall, our hierarchical few-shot model is better at identifying a wider range of instrument classes than the baseline. This is important if we desire to make systems that are more robust to biases in the training data and, thus, can classify more a diverse set of instrument types.

\vspace{-0.25cm}

\section{Conclusion}
We presented an approach for incorporating hierarchical structures in a few-shot learning model for the purpose of improving classification performance on classes outside of the training distribution. Our method builds on top of prototypical networks by computing prototypical representations at fine and coarse granularities, as defined by a class hierarchy. We showed that our proposed method yields statistically significant increases in classification performance and significant decreases mistake severity when evaluated on a classification task composed of unseen musical instruments. Moreover, we found that the choice of hierarchical structure is not arbitrary, and using a hierarchy based on the sound production mechanisms of musical instruments had the best results. We hope our work enables users with diverse cultural backgrounds with the ability to classify diverse collections of musical instruments. Future directions include examining new types of hierarchies, learning multiple hierarchies simultaneously, and the unsupervised discovery of hierarchies from unlabeled data.

\section{Acknowledgements}
This work was funded, in part, by USA National Science Foundation Award 1901456. 

\newpage
\bibliography{ISMIRtemplate}

\begin{thebibliography}{10}
\providecommand{\url}[1]{#1}
\csname url@samestyle\endcsname
\providecommand{\newblock}{\relax}
\providecommand{\bibinfo}[2]{#2}
\providecommand{\BIBentrySTDinterwordspacing}{\spaceskip=0pt\relax}
\providecommand{\BIBentryALTinterwordstretchfactor}{4}
\providecommand{\BIBentryALTinterwordspacing}{\spaceskip=\fontdimen2\font plus
\BIBentryALTinterwordstretchfactor\fontdimen3\font minus
  \fontdimen4\font\relax}
\providecommand{\BIBforeignlanguage}[2]{{%
\expandafter\ifx\csname l@#1\endcsname\relax
\typeout{** WARNING: IEEEtran.bst: No hyphenation pattern has been}%
\typeout{** loaded for the language `#1'. Using the pattern for}%
\typeout{** the default language instead.}%
\else
\language=\csname l@#1\endcsname
\fi
#2}}
\providecommand{\BIBdecl}{\relax}
\BIBdecl

\bibitem{fu2010survey}
Z.~Fu, G.~Lu, K.~M. Ting, and D.~Zhang, ``A survey of audio-based music
  classification and annotation,'' \emph{IEEE Transactions on Multimedia},
  vol.~13, no.~2, pp. 303--319, 2010.

\bibitem{eronen2000musical}
A.~Eronen and A.~Klapuri, ``Musical instrument recognition using cepstral
  coefficients and temporal features,'' in \emph{2000 IEEE International
  Conference on Acoustics, Speech, and Signal Processing. Proceedings (Cat. No.
  00CH37100)}, vol.~2.\hskip 1em plus 0.5em minus 0.4em\relax IEEE, 2000, pp.
  II753--II756.

\bibitem{krishna2004music}
A.~Krishna and T.~V. Sreenivas, ``Music instrument recognition: from isolated
  notes to solo phrases,'' in \emph{2004 IEEE International Conference on
  Acoustics, Speech, and Signal Processing}, vol.~4.\hskip 1em plus 0.5em minus
  0.4em\relax IEEE, 2004, pp. iv--iv.

\bibitem{mesaros2016metrics}
A.~Mesaros, T.~Heittola, and T.~Virtanen, ``Metrics for polyphonic sound event
  detection,'' \emph{Applied Sciences}, vol.~6, no.~6, p. 162, 2016.

\bibitem{mesaros2016tut}
------, ``Tut database for acoustic scene classification and sound event
  detection,'' in \emph{2016 24th European Signal Processing Conference
  (EUSIPCO)}.\hskip 1em plus 0.5em minus 0.4em\relax IEEE, 2016, pp.
  1128--1132.

\bibitem{salamon2017deep}
J.~Salamon and J.~P. Bello, ``Deep convolutional neural networks and data
  augmentation for environmental sound classification,'' \emph{IEEE Signal
  Processing Letters}, vol.~24, no.~3, pp. 279--283, 2017.

\bibitem{savage2011art}
S.~Savage, \emph{The art of digital audio recording: A practical guide for home
  and studio}.\hskip 1em plus 0.5em minus 0.4em\relax Oxford University Press,
  2011.

\bibitem{owsinski2013mixing}
B.~Owsinski, \emph{The mixing engineer's handbook}.\hskip 1em plus 0.5em minus
  0.4em\relax Nelson Education, 2013.

\bibitem{saha-vision-2020}
A.~Saha and A.~M. Piper, ``Understanding audio production practices of people
  with vision impairments,'' in \emph{The 22nd International ACM SIGACCESS
  Conference on Computers and Accessibility (ASSETS ’20), October 26–28,
  2020, Virtual Event, Greece.}\hskip 1em plus 0.5em minus 0.4em\relax IEEE,
  2020.

\bibitem{tanaka2016haptic}
A.~Tanaka and A.~Parkinson, ``Haptic wave: A cross-modal interface for visually
  impaired audio producers,'' in \emph{Proceedings of the 2016 CHI Conference
  on Human Factors in Computing Systems}, 2016, pp. 2150--2161.

\bibitem{bosch2012comparison}
J.~J. Bosch, J.~Janer, F.~Fuhrmann, and P.~Herrera, ``A comparison of sound
  segregation techniques for predominant instrument recognition in musical
  audio signals.'' in \emph{International Society for Music Information
  Retrieval (ISMIR) Conference}.\hskip 1em plus 0.5em minus 0.4em\relax
  Citeseer, 2012, pp. 559--564.

\bibitem{humphrey-openmic-2018}
E.~Humphrey, S.~Durand, and B.~McFee, ``Openmic-2018: An open data-set for
  multiple instrument recognition.'' in \emph{International Society for Music
  Information Retrieval (ISMIR) Conference}, Paris, France, 2018, pp. 438--444.

\bibitem{hung-timbre-2018}
Y.-N. Hung and Y.~Yang, ``Frame-level instrument recognition by timbre and
  pitch,'' in \emph{{Proceedings of the 20th International Society for Music
  Information Retrieval Conference}}.\hskip 1em plus 0.5em minus 0.4em\relax
  Paris, France: ISMIR, 2018.

\bibitem{gururani-attention-2019}
S.~Gururani, M.~Sharma, and A.~Lerch, ``An attention mechanism for musical
  instrument recognition,'' in \emph{International Society for Music
  Information Retrieval (ISMIR) Conference}, 2019.

\bibitem{hung-multitask-2019}
Y.~{Hung}, Y.~{Chen}, and Y.~{Yang}, ``Multitask learning for frame-level
  instrument recognition,'' in \emph{ICASSP 2019 - 2019 IEEE International
  Conference on Acoustics, Speech and Signal Processing (ICASSP)}, 2019, pp.
  381--385.

\bibitem{taenzer2019investigating}
M.~Taenzer, J.~Abe{\ss}er, S.~I. Mimilakis, C.~Wei{\ss}, M.~M{\"u}ller, and
  H.~Lukashevich, ``Investigating cnn-based instrument family recognition for
  western classical music recordings.'' in \emph{Proceedings of the 20th
  International Society for Music Information Retrieval Conference}.\hskip 1em
  plus 0.5em minus 0.4em\relax ISMIR, 2019, pp. 612--619.

\bibitem{kratimenos2021augmentation}
A.~Kratimenos, K.~Avramidis, C.~Garoufis, A.~Zlatintsi, and P.~Maragos,
  ``Augmentation methods on monophonic audio for instrument classification in
  polyphonic music,'' in \emph{2020 28th European Signal Processing Conference
  (EUSIPCO)}.\hskip 1em plus 0.5em minus 0.4em\relax IEEE, 2021, pp. 156--160.

\bibitem{lostanlen2018extended}
V.~Lostanlen, J.~And{\'e}n, and M.~Lagrange, ``Extended playing techniques: the
  next milestone in musical instrument recognition,'' in \emph{Proceedings of
  the 5th International Conference on Digital Libraries for Musicology}, 2018,
  pp. 1--10.

\bibitem{kim-ised-2017}
B.~Kim and B.~Pardo, ``I-sed: An interactive sound event detector,'' in
  \emph{Proceedings of the 22nd International Conference on Intelligent User
  Interfaces}, ser. IUI '17.\hskip 1em plus 0.5em minus 0.4em\relax New York,
  NY, USA: Association for Computing Machinery, 2017, p. 553–557.

\bibitem{cartwright2019crowdsourcing}
M.~Cartwright, G.~Dove, A.~E. M{\'e}ndez~M{\'e}ndez, J.~P. Bello, and O.~Nov,
  ``Crowdsourcing multi-label audio annotation tasks with citizen scientists,''
  in \emph{Proceedings of the 2019 CHI Conference on Human Factors in Computing
  Systems}, 2019, pp. 1--11.

\bibitem{hornbostel-classification-1961}
E.~M. von Hornbostel and C.~Sachs, ``Classification of musical instruments:
  Translated from the original german by anthony baines and klaus p.
  wachsmann,'' \emph{The Galpin Society Journal}, vol.~14, pp. 3--29, 1961.

\bibitem{kartomi1990concepts}
M.~J. Kartomi, \emph{On concepts and classifications of musical
  instruments}.\hskip 1em plus 0.5em minus 0.4em\relax University of Chicago
  Press Chicago, 1990.

\bibitem{snell-prototypical-2017}
J.~Snell, K.~Swersky, and R.~Zemel, ``Prototypical networks for few-shot
  learning,'' in \emph{Advances in Neural Information Processing Systems},
  I.~Guyon, U.~V. Luxburg, S.~Bengio, H.~Wallach, R.~Fergus, S.~Vishwanathan,
  and R.~Garnett, Eds., vol.~30.\hskip 1em plus 0.5em minus 0.4em\relax Curran
  Associates, Inc., 2017.

\bibitem{oded2006datamining}
O.~Maimon and R.~Lior, \emph{Data Mining and Knowledge Discovery
  Handbook}.\hskip 1em plus 0.5em minus 0.4em\relax Springer, 2006, ch.
  Clustering methods.

\bibitem{wang-fewshotsed-2020}
Y.~{Wang}, J.~{Salamon}, N.~J. {Bryan}, and J.~{Pablo Bello}, ``Few-shot sound
  event detection,'' in \emph{ICASSP 2020 - 2020 IEEE International Conference
  on Acoustics, Speech and Signal Processing (ICASSP)}, 2020, pp. 81--85.

\bibitem{benetos-nmf-2006}
E.~{Benetos}, M.~{Kotti}, and C.~{Kotropoulos}, ``Musical instrument
  classification using non-negative matrix factorization algorithms and subset
  feature selection,'' in \emph{2006 IEEE International Conference on Acoustics
  Speech and Signal Processing Proceedings}, vol.~5, 2006, pp. V--V.

\bibitem{eronen-mfcc-2000}
A.~{Eronen} and A.~{Klapuri}, ``Musical instrument recognition using cepstral
  coefficients and temporal features,'' in \emph{2000 IEEE International
  Conference on Acoustics, Speech, and Signal Processing. Proceedings (Cat.
  No.00CH37100)}, vol.~2, 2000, pp. II753--II756 vol.2.

\bibitem{lostanlen-spiral-2016}
V.~Lostanlen and C.-E. Cella, ``Deep convolutional networks on the pitch spiral
  for music instrument recognition,'' in \emph{International Society for Music
  Information Retrieval (ISMIR) Conference}, 2016.

\bibitem{essid-hierarchicalsolos-2006}
S.~{Essid}, G.~{Richard}, and B.~{David}, ``Hierarchical classification of
  musical instruments on solo recordings,'' in \emph{2006 IEEE International
  Conference on Acoustics Speech and Signal Processing Proceedings}, vol.~5,
  2006, pp. V--V.

\bibitem{han-predominant-2017}
Y.~Han, J.~Kim, K.~Lee, Y.~Han, J.~Kim, and K.~Lee, ``Deep convolutional neural
  networks for predominant instrument recognition in polyphonic music,''
  \emph{IEEE/ACM Trans. Audio, Speech and Lang. Proc.}, vol.~25, no.~1, p.
  208–221, Jan. 2017.

\bibitem{gururani-iad-2018}
S.~Gururani, C.~Summers, and A.~Lerch, ``Instrument activity detection in
  polyphonic music using deep neural networks,'' in \emph{{Proceedings of the
  20th International Society for Music Information Retrieval
  Conference}}.\hskip 1em plus 0.5em minus 0.4em\relax Paris, France: ISMIR,
  2018.

\bibitem{stein-hierarchicaltext-2019}
R.~A. Stein, P.~A. Jaques, and J.~F. Valiati, ``An analysis of hierarchical
  text classification using word embeddings,'' \emph{Information Sciences},
  vol. 471, pp. 216--232, 2019.

\bibitem{ankit-image_hierarchical-2020}
A.~Dhall, A.~Makarova, O.~Ganea, D.~Pavllo, M.~Greeff, and A.~Krause,
  ``Hierarchical image classification using entailment cone embeddings,'' in
  \emph{Proceedings of the IEEE/CVF Conference on Computer Vision and Pattern
  Recognition Workshops}, 2020, pp. 836--837.

\bibitem{sun-hierarchicalimage-2019}
S.~Sun, Q.~Sun, K.~Zhou, and T.~Lv, ``Hierarchical attention prototypical
  networks for few-shot text classification,'' in \emph{Proceedings of the 2019
  Conference on Empirical Methods in Natural Language Processing and the 9th
  International Joint Conference on Natural Language Processing
  (EMNLP-IJCNLP)}.\hskip 1em plus 0.5em minus 0.4em\relax Hong Kong, China:
  Association for Computational Linguistics, Nov. 2019, pp. 476--485.

\bibitem{bertinetto-mistakes-2020}
L.~Bertinetto, R.~Mueller, K.~Tertikas, S.~Samangooei, and N.~A. Lord, ``Making
  better mistakes: Leveraging class hierarchies with deep networks,'' in
  \emph{Proceedings of the IEEE/CVF Conference on Computer Vision and Pattern
  Recognition}, 2020, pp. 12\,506--12\,515.

\bibitem{essid-2005-instrument}
S.~Essid, G.~Richard, and B.~David, ``Instrument recognition in polyphonic
  music based on automatic taxonomies,'' \emph{IEEE Transactions on Audio,
  Speech, and Language Processing}, vol.~14, no.~1, pp. 68--80, 2005.

\bibitem{kitahara-hierarchicalnonregistered-2004}
T.~{Kitahara}, M.~{Goto}, and H.~G. {Okuno}, ``Category-level identification of
  non-registered musical instrument sounds,'' in \emph{2004 IEEE International
  Conference on Acoustics, Speech, and Signal Processing}, vol.~4, 2004, pp.
  iv--iv.

\bibitem{fernandez-hierarchicalsequencernn-2007}
S.~Fern{\'a}ndez, A.~Graves, and J.~Schmidhuber, ``Sequence labelling in
  structured domains with hierarchical recurrent neural networks,'' in
  \emph{Proceedings of the 20th International Joint Conference on Artificial
  Intelligence, IJCAI 2007}, 2007.

\bibitem{sanabria-hierarchicalctc-2018}
R.~Sanabria and F.~Metze, ``Hierarchical multitask learning with ctc,'' in
  \emph{2018 IEEE Spoken Language Technology Workshop (SLT)}, 2018, pp.
  485--490.

\bibitem{krishna-hierarchicalspeech-2018}
K.~Krishna, S.~Toshniwal, and K.~Livescu, ``Hierarchical multitask learning for
  ctc-based speech recognition,'' \emph{arXiv preprint arXiv:1807.06234}, 2018.

\bibitem{manilow-hierarchical-2020}
E.~Manilow, G.~Wichern, and J.~Le~Roux, ``Hierarchical musical instrument
  separation,'' in \emph{International Society for Music Information Retrieval
  (ISMIR) Conference}, Oct. 2020, pp. 376--383.

\bibitem{xu2016hierarchical}
Y.~Xu, Q.~Huang, W.~Wang, and M.~D. Plumbley, ``Hierarchical learning for
  dnn-based acoustic scene classification,'' \emph{arXiv preprint
  arXiv:1607.03682}, 2016.

\bibitem{jati-hierarchical_loss-2019}
A.~{Jati}, N.~{Kumar}, R.~{Chen}, and P.~{Georgiou}, ``Hierarchy-aware loss
  function on a tree structured label space for audio event detection,'' in
  \emph{ICASSP 2019 - 2019 IEEE International Conference on Acoustics, Speech
  and Signal Processing (ICASSP)}, 2019, pp. 6--10.

\bibitem{cramer-taxonet-2020}
J.~Cramer, V.~Lostanlen, A.~Farnsworth, J.~Salamon, and J.~P. Bello, ``Chirping
  up the right tree: Incorporating biological taxonomies into deep bioacoustic
  classifiers,'' in \emph{ICASSP 2020-2020 IEEE International Conference on
  Acoustics, Speech and Signal Processing (ICASSP)}.\hskip 1em plus 0.5em minus
  0.4em\relax IEEE, 2020, pp. 901--905.

\bibitem{cheng-fewshotsed-2019}
K.~{Cheng}, S.~{Chou}, and Y.~{Yang}, ``Multi-label few-shot learning for sound
  event recognition,'' in \emph{2019 IEEE 21st International Workshop on
  Multimedia Signal Processing (MMSP)}, 2019, pp. 1--5.

\bibitem{shi-fewshotacoustic-2020}
B.~Shi, M.~Sun, K.~C. Puvvada, C.-C. Kao, S.~Matsoukas, and C.~Wang, ``Few-shot
  acoustic event detection via meta learning,'' in \emph{ICASSP 2020 - 2020
  IEEE International Conference on Acoustics, Speech and Signal Processing
  (ICASSP)}, 2020, pp. 76--80.

\bibitem{wang-fewshotdrum-2020}
Y.~Wang, J.~Salamon, M.~Cartwright, N.~J. Bryan, and J.~P. Bello, ``Few-shot
  drum transcription in polyphonic music,'' in \emph{International Society for
  Music Information Retrieval (ISMIR) Conference}, 2020.

\bibitem{bittner-medleydb-2014}
R.~Bittner, J.~Salamon, M.~Tierney, M.~Mauch, C.~Cannam, and J.~Bello,
  ``Medleydb: A multitrack dataset for annotation-intensive mir research,'' in
  \emph{15th International Society for Music Information Retrieval Conference
  (ISMIR)}, 2014.

\bibitem{bittner-medleydb2-2016}
\BIBentryALTinterwordspacing
R.~Bittner, J.~Wilkins, H.~Yip, and J.~P. Bello, ``Medleydb 2.0 audio,'' Aug.
  2016. [Online]. Available: \url{https://doi.org/10.5281/zenodo.1715175}
\BIBentrySTDinterwordspacing

\bibitem{bittner-pysox-2016}
R.~Bittner, E.~Humphrey, and J.~Bello, ``Pysox: Leveraging the audio signal
  processing power of sox in python,'' in \emph{Proceedings of the
  International Society for Music Information Retrieval Conference Late
  Breaking and Demo Papers}, 2016.

\end{thebibliography}

\end{document}